\documentclass{aa}  

\usepackage{graphicx}
\usepackage{color}
\usepackage{txfonts}
\usepackage{ulem}
\begin{document} 

\makeatletter
\def\instrefs#1{{\def\scsep{\def\scsep{,}}\@for\w:=#1\do{\scsep\ref{inst:\w}}}}
\renewcommand{\inst}[1]{\unskip$^{\instrefs{#1}}$}

\title{The strange case of \ion{Na}{i} in the atmosphere of HD209458~b:}
\subtitle{Reconciling low- and high-resolution spectroscopic observations}

\author{G. Morello\inst{iac,ull,inaf}
\and
N. Casasayas-Barris\inst{lo}
\and
J. Orell-Miquel\inst{iac,ull}
\and
E. Pall{\'e}\inst{iac,ull}
\and
G. Cracchiolo\inst{inaf,unipa}
\and
G. Micela\inst{inaf}
}

\institute{
\label{inst:iac}Instituto de Astrof\'isica de Canarias (IAC), 38205 La Laguna, Tenerife, Spain
\and 
\label{inst:ull}Departamento de Astrof\'isica, Universidad de La Laguna (ULL), 38206, La Laguna, Tenerife, Spain
\and
\label{inst:lo}Leiden Observatory, Leiden University, Postbus 9513, 2300 RA Leiden, The Netherlands
\and
\label{inst:inaf}INAF- Palermo Astronomical Observatory, Piazza del Parlamento, 1, 90134 Palermo, Italy
\and
\label{inst:unipa}University of Palermo, Department of Physics and Chemistry ``Emilio Segr{\`e}''
}


 
  \abstract
   {}
   {We aim to investigate the origin of the discrepant results reported in the literature about the presence of \ion{Na}{i} in the atmosphere of HD209458~b, based on low- and high-resolution transmission spectroscopy.}
   {We generated synthetic planetary atmosphere models and we compared them with the transmission light curves and spectra observed in previous studies. Our models account for the stellar limb-darkening and Rossiter-McLaughlin (RM) effects, and contemplate various possible scenarios for the planetary atmosphere.}
   {We reconciled the discrepant results by identifying a range of planetary atmospheres that are consistent with previous low- and high-resolution spectroscopic observations. Either both datasets are interpreted as consistent with a total absence of \ion{Na}{i} in the planetary atmosphere (with Hubble Space Telescope data being affected by limb darkening), or the terminator temperature of HD209458~b has to have an upper limit of about 1000K. In particular, we find that 1D transmission spectra with lower-than-equilibrium temperatures can also explain the previously reported detection of absorption signal at low resolution due to differential transit depth in adjacent bands, while the cores of the \ion{Na}{i} D lines may be masked by the strong RM signal seen at high resolution. We also rule out high-altitude clouds, which would otherwise mask the absorption signal at low resolution, as the source of the discrepancies.}
   {This work highlights the synergies between different observing techniques, specifically low- and high-resolution spectroscopy, to fully characterise transiting exoplanet systems.}


\maketitle
%

\section{Introduction}

The characterisation of exoplanet atmospheres relies on precise measurements of the spectroscopic signal that they imprint on the observed starlight. For a planet transiting in front of its host star, the atmosphere acts as a wavelength-dependent annulus that affects the fraction of occulted stellar flux \citep{brown2001}.
If an atom or molecule is present in the planetary atmosphere, it causes extra absorption at specific wavelengths corresponding to electronic transition lines.
The transit depth is defined as the apparent planet-to-star area ratio, $p^2 = (R_p/R_*)^2$, $R_p$ and $R_*$ being the planet and star radii. Differences in transit depths obtained on multiple passbands or wavelength bins have led to the detection of many chemical species in the atmospheres of exoplanets (e.g. \citealp{charbonneau2002,vidal-madjar2003,fossati2010,deming2013,damiano2017}). High-resolution spectroscopy has enabled us to resolve the atmospheric spectral features and to trace their Doppler shift as it varies with the orbital phase (e.g. \citealp{snellen2010,casasayas-barris2017,casasayas-barris2018,casasayas-barris2019,palle2020,stangret2020}).
Typically, the signal from the planet atmosphere is of the order of 10$^{-4}$ or less, relatively to the host star flux. It is therefore necessary to disentangle this small signal from other contaminating effects with similar amplitudes, such as stellar limb darkening \citep{howarth2011,csizmadia2013,morello2017,yan2017}, magnetic activity \citep{ballerini2012,oshagh2014,cracchiolo2021a,cracchiolo2021b}, and planet self- and phase-blend effects \citep{kipping2010,martin-lagarde2020,morello2021}.

HD209458~b is the first planet with a reported detection of a chemical species in its atmosphere, namely \ion{Na}{i} \citep{charbonneau2002}. The observations were performed with the Hubble Space Telescope (HST)/Space Telescope Imaging Spectrograph (STIS) using the G750M filter. In particular, \cite{charbonneau2002} inferred \ion{Na}{i} absorption from the larger transit depth on a narrow band centred on the resonance doublet at 5893 $\AA$ relative to adjacent bands. \cite{sing2008} reanalysed the same data, confirming the previous \ion{Na}{i} detection and resolving the doublet lines. Other HST/STIS datasets led to the possible detection of \ion{H}{i}, \ion{O}{i}, \ion{C}{ii} \citep{vidal-madjar2004}, \ion{Mg}{i} \citep{vidal-madjar2013}, and \ion{Fe}{ii} \citep{cubillos2020}. The near-infrared spectrum taken with the HST/Wide Field Camera 3 (WFC3) also revealed H$_2$O absorption \citep{deming2013,tsiaras2016}.
Ground-based observations provided independent confirmation of the \ion{Na}{i} feature with higher spectral resolution \citep{snellen2008} and also revealed other features attributed to \ion{He}{i} \citep{alonso-floriano2019}, \ion{Ca}{i}, \ion{Sc}{ii}, \ion{H}{i} \citep{astudillo-defru2013}, CO \citep{snellen2010}, H$_2$O \citep{sanchez-lopez2019}, HCN, CH$_4$, C$_2$H$_2$ and NH$_3$ \citep{giacobbe2021}.
However, recent studies cast doubts on the \ion{Na}{i} detection as well as that of other atomic and ionic species \citep{casasayas-barris2020,casasayas-barris2021}. In particular, the transmission spectra observed with the High Accuracy Radial velocity Planet Searcher for the Northern hemisphere (HARPS-N), mounted on the Telescopio Nazionale Galileo (TNG) at the Observatory of Roque de los Muchachos (ORM) in Spain, and the Echelle Spectrograph for Rocky Exoplanet and Stable Spectroscopic Observations (ESPRESSO), mounted on the Very Large Telescope (VLT) at the European Southern Observatory (ESO) of Cerro Paranal in Chile, reveal several features due to the Rossiter-McLaughlin (RM) effect and no evidence of absorbing species in the exoplanet atmosphere. 

In this paper, we try to reconcile these apparently conflicting results that have appeared using different instruments, observing techniques, and analysis methods. Section \ref{sec:hst} presents a reanalysis of the HST/STIS observations that led to the first announcement of \ion{Na}{i} in the atmosphere of HD209458~b with updated data detrending techniques, transit modelling tools, and system parameters. Section \ref{sec:simulations} describes our models of the planetary atmosphere. From them, we extract time series and spectra following the same procedures adopted by previously published studies of low and high spectral resolution observations aimed at exoplanet atmospheric characterisation, along with other stellar and planetary effects. Section \ref{sec:discussion} discusses the comparison between our simulations and the observations, providing a range of planetary atmosphere scenarios that could explain the apparently discrepant results. Section \ref{sec:conclusions} summarises the conclusions of our study. 

\section{Reanalysis of HST data}
\label{sec:hst}

\begin{table}
\caption{Summary of HST observations for identification in the online archives. More technical information is reported in Section \ref{sec:hst_obs}. }
\label{tab:hst_obs_id}
\centering
\begin{tabular}{ccc}
\hline\hline
Root name & Number of spectra\tablefootmark{a}  & UT start date \\
\hline
o63302030 & 36, 36, 36, 35 & 2000-04-28 \\
o63303030 & 36, 36, 36, 35 & 2000-05-05 \\
o63304030 & 36, 36, 36, 35 & 2000-05-12 \\
\hline
\end{tabular}
\tablefoot{\tablefoottext{a}{For each of the four orbits.}}
\end{table}

\begin{table*}
\caption{Adopted HD209458 system parameters.}
\label{tab:sys_params}
\centering
\begin{tabular}{lcc}
\hline\hline
\multicolumn{3}{c}{Stellar parameters} \\
\hline
T$_{*,\mathrm{eff}}$ [K] & 6065$\pm$50 & \cite{torres2008} \\
$\log{g_*}$ [cgs] & 4.361$\pm$0.008 & " \\
$[\mathrm{Fe/H}]_*$ [dex] & 0.00$\pm$0.05 & " \\
$M_*$ [$M_{\odot}$] & 1.119$\pm$0.033 & " \\
$R_*$ [$R_{\odot}$] & 1.155$\pm$0.016 & " \\
\hline
\multicolumn{3}{c}{Planetary parameters} \\
\hline
$M_{\mathrm{p}}$ [$M_{\mathrm{Jup}}$] & 0.685$\pm$0.015 & " \\
$R_{\mathrm{p}}$ [$R_{\mathrm{Jup}}$] & 1.359$\pm$0.019 & " \\
$a$ [au] & 0.04707$\pm$0.00047 & " \\
\hline
\multicolumn{3}{c}{Transit parameters} \\
\hline
$P$ [day] & 3.52474859$\pm$0.00000038 & \cite{knutson2007} \\
$E.T.$ [HJD] & 2452826.628521$\pm$0.000087 & " \\
$b$ [R$_*$] & 0.516$\pm$0.006 & \cite{morello2018} \\
$T_0$ [s] & 9500$\pm$6 & " \\
\hline
\multicolumn{3}{c}{RM parameters} \\
\hline
$K_{\mathrm{p}}$ [km/s] & 145.0$\pm$1.6 & \cite{casasayas-barris2021} \\
$\lambda$ [deg] & 1.58$\pm$0.08 & " \\
$v \sin{i_*}$ [km/s] & 4.228$\pm$0.007 & " \\
\hline
\end{tabular}
\end{table*}

\subsection{Observations}
\label{sec:hst_obs}

We reanalysed three transits of HD209458~b observed with HST/STIS (GO-8789, PI: Brown) using the G750M grism, taken on 28 April, 5 May and 5 December 2000. Table \ref{tab:hst_obs_id} provides their identifiers. The spectra cover the 5808-6380 $\AA$ wavelength range and have a resolving power of 5540 at the central wavelength of 6094 $\AA$. Each visit contains 143 spectra distributed over four HST orbits: one before, two during, and one after the transit event. The four HST orbits were preceded by another one to enable instrumental settling, but it was not included in the scientific data analysis. The integration time is 60 s per frame, followed by a reset time of about 20 s. The interval between consecutive HST orbits is about 50 min. 
The spectral trace forms an angle of less than 0.5$^{\circ}$ with the longest side of the 64$\times$1024 pixel detector, and it is stable within the same pixel rows during each visit.

\subsection{Data analysis}
\subsubsection{Extraction}
\label{sec:hst_extraction}
We downloaded the flat-fielded science images (extension: flt.fits) from the Mikulski Archive for Space Telescopes (MAST)\footnote{\url{https://archive.stsci.edu}}. We considered a rectangular aperture of 17$\times$1024 pixels with the trace at its centre and summed along the columns to extract the 1D spectra. As a starting point, we adopted the wavelength solution from the corresponding archive spectra (extension: x1d.fits) and then computed the cross-correlations with a template spectrum to align the extracted spectra in the stellar rest frame. The template spectrum was calculated as described in Section \ref{sec:sim_star}, and purposely degraded to the same resolution as the observations.

Our analysis focused on a portion of the spectrum containing the \ion{Na}{i} doublet. We extracted the flux time series for three bands identical to the `narrow' ones selected by \cite{charbonneau2002} using two different sets of apertures. The first set consisted of rectangular apertures, similar to the ones used for the wavelength calibration, but limited in the dispersion axis according to the bands definition. In this case, the flux is the simple sum over the pixels contained in the rectangular aperture. The second set consisted of tilted rectangular apertures with sides parallel and perpendicular to the spectral trace, which was previously calculated by a linear fit on the centroids resulting from Gaussian fits on the detector columns. When calculating the flux from the fractions of pixels within the tilted apertures we accounted for the non-uniform distribution inside the pixel, that was approximated by a linear function in the cross-dispersion direction.
The resulting light curves have almost identical shapes, regardless of the extraction method, but the tilted apertures led to higher fluxes by $\sim$0.01$\%$, $\sim$0.04$\%$ and $\sim$0.10$\%$ for the left, central, and right bands, respectively. This behaviour is expected as the small adjustment for the orientation of the spectral trace reduces the flux losses from the tails of the point spread function. For the rest of the analysis described in this paper, both methods led to indistinguishable results.

\subsubsection{Detrending}
\label{sec:hst_detrending}
The HST time series exhibit well-known systematic effects, usually referred to as short- and long-term ramps \citep{brown2001b}. The short-term ramp is a flux variation that follows a highly repeatable pattern for each orbit of the same visit, excluding the first orbit. The long-term ramp approximates a linear trend in the transit timescale. We adopted the \texttt{divide-oot} method to correct for the ramp effects \citep{berta2012}. Following this method, the two in-transit orbits are divided by the time-weighted mean of the two out-of-transit orbits. Since the last orbit contains one less spectrum than the others, we duplicated the last point to enable the operations described above. We note that the added point falls into the plateau at the end of the ramp and out of transit, so it is expected to be similar to the previous one.
\begin{figure*}
\centering
\includegraphics[width=0.48\hsize]{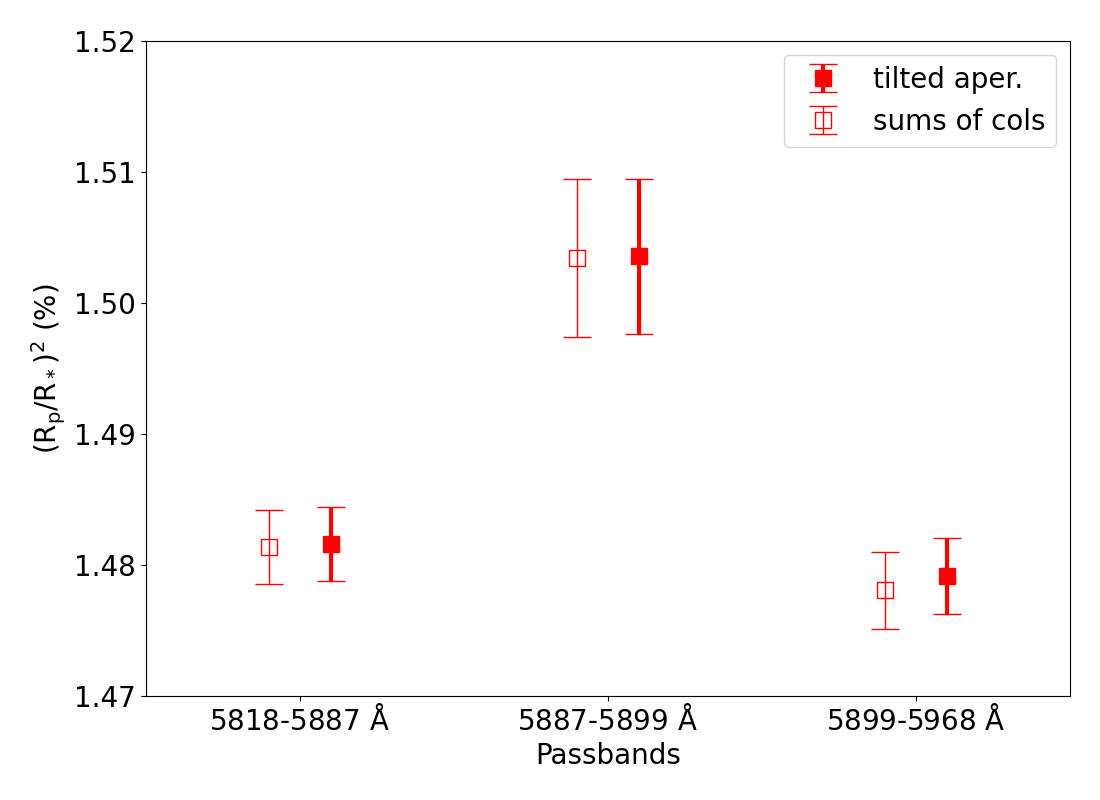}
\includegraphics[width=0.51\hsize]{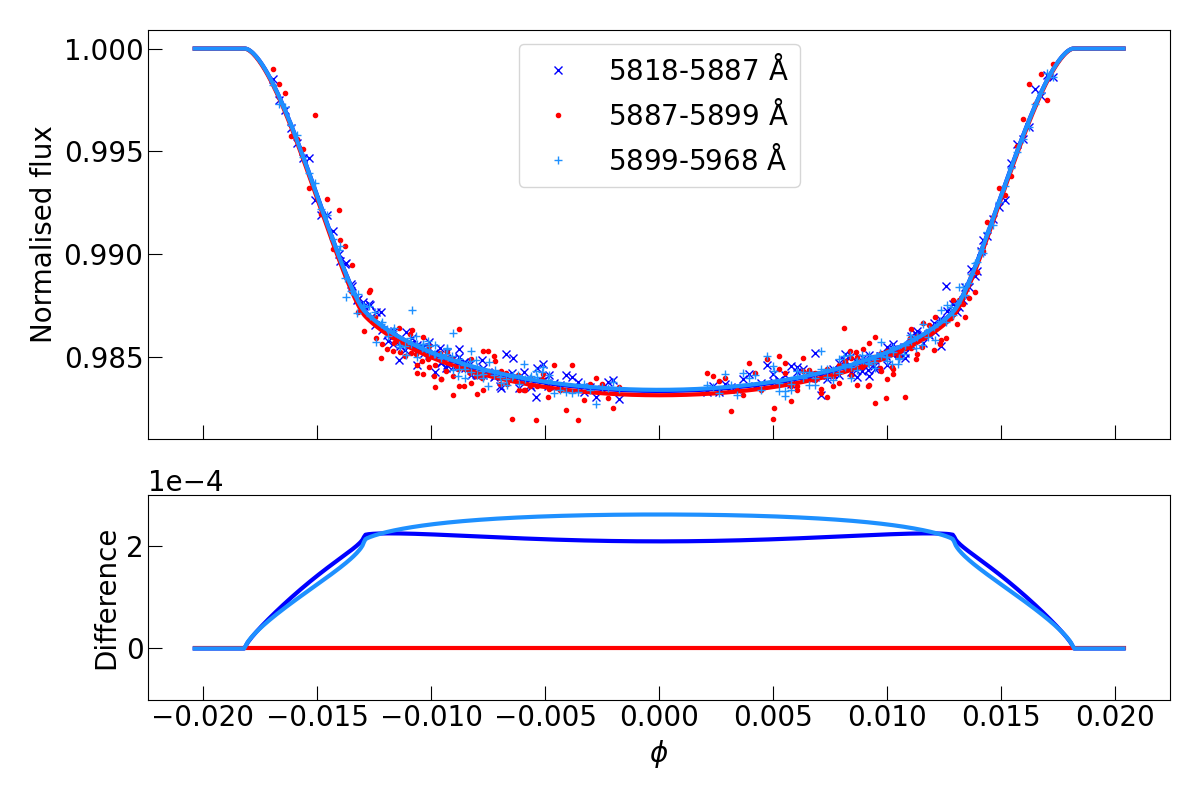}
\caption{Left: Best-fit transit depths to HST data around the Na I lines, as reported in Table \ref{tab:hst_tdepths}. Right: \texttt{divide-oot} detrended and phase-folded HST light curves (see Section \ref{sec:hst_detrending}) for the 5818-5887 $\AA$ (blue crosses), 5887-5899 $\AA$ (red dots), and 5899-5968 $\AA$ (dodger blue pluses) passbands, along with best-fit transit models (continuous coloured lines). The detrended out-of-transit data are not represented, as they are identical to 1 by definition of the \texttt{divide-oot} method \citep{berta2012}. The bottom panel shows the transit model differences with respect to that of the central band. The slightly different shape between the two side bands is due to differential limb darkening.}
\label{fig:tspec_cha02narrow}
\end{figure*}

\subsubsection{Fitting}
We combined the detrended orbits from the three visits into phase-folded time series to perform joint light-curve fits.
The transit models were computed with \texttt{PYLIGHTCURVE}\footnote{\url{https://github.com/ucl-exoplanets/pylightcurve}} \citep{tsiaras2016}. Since we were primarily interested in the differences in transit depth between the selected passbands, we kept the planet-to-star radii ratio ($p$) as the only free parameter. Table \ref{tab:sys_params} reports the system parameters along with the transit parameters that were fixed in the light-curve fits. The stellar limb-darkening coefficients were fixed to the values obtained with \texttt{ExoTETHyS}\footnote{\url{https://github.com/ucl-exoplanets/exotethys}} \citep{morello2020joss,morello2020}) using the four-coefficient law \citep{claret2000} and the \texttt{STAGGER} grid of stellar spectra \citep{magic2015,chiavassa2018}.
The fitting procedure included a preliminary least-squares minimisation using \texttt{scipy.optimize.minimize} with the Nelder-Mead method \citep{nelder1965}. Then we ran \texttt{emcee}\footnote{\url{https://github.com/dfm/emcee}} \citep{emceev3joss} with 80 walkers and 200\,000 iterations. Each walker was initialised with a random value close to the least-squares parameter estimate. The first 100\,000 iterations were discarded as burn-in. The remaining samples were used to determine the final best-fit parameters (median) and error bars (absolute differences between the 16th and 84th percentiles and the median).

\subsection{Results}
Table \ref{tab:hst_tdepths} reports the measured transit depths for the three selected passbands with the two extraction methods described in Section \ref{sec:hst_extraction}. Figure \ref{fig:tspec_cha02narrow} (left panel) shows the same results. The two sets of results are nearly identical with 1-3$\%$ smaller error bars when using tilted apertures. They indicate a greater transit depth in the central band containing the \ion{Na}{i} doublet, and no significant difference in transit depth between the two side bands. We calculated the \ion{Na}{i} absorption signal as the difference between the transit depth in the central band and the average of the two side values, obtaining 232$\pm$62 ppm (tilted apertures) and 237$\pm$64 ppm (sum of pixel columns). These results confirm that \ion{Na}{i} is detected in the atmosphere of HD209458~b at 3.7$\sigma$.

We note that the differences between the best-fit light-curve models for the side bands and for the central band are overall positive during transit, and show a plateau at $>$200 ppm at orbital phases $| \Delta \phi | <$0.013 (see bottom right panel of Figure \ref{fig:tspec_cha02narrow}). If stellar limb-darkening was the only wavelength-dependent effect, the light-curve differences should have a typical double-horned modulation with both positive and negative values \citep{rosenblatt1971,tingley2004,parviainen2019}. The numerical closeness between the plateau in the light-curve differences and the measured \ion{Na}{i} absorption feature also suggests that the results of the analysis are not strongly influenced by the adopted stellar limb-darkening model.
However, some authors pointed out that stellar limb-darkening models may be biased, because of unaccounted-for effects and lack of empirical validation \citep{csizmadia2013,espinoza2015,morello2018}.
We also attempted light-curve fitting with free limb-darkening coefficients. The results point towards a smaller (non-significant) \ion{Na}{i} absorption feature and enhanced differential limb darkening, but they are inconclusive because of the low signal-to-noise ratio (S/N) of the data and strong parameter degeneracies. Better quality data are needed to conclude as to the authenticity of the \ion{Na}{i} detection at low resolution.

\begin{table}
\caption{Best-fit transit depths to HST data and \ion{Na}{i} absorption signal.}
\label{tab:hst_tdepths}
\centering
\begin{tabular}{ccc}
\hline\hline
Passband & \multicolumn{2}{c}{Transit Depth} \\
($\mathrm{\AA}$) & tilted aper. ($\%$) & sums of col. ($\%$) \\
\hline
nl: 5818-5887 & 1.482$\pm$0.028 & 1.481$\pm$0.029 \\
nc: 5887-5899 & 1.504$\pm$0.059 & 1.503$\pm$0.060 \\
nr: 5899-5968 & 1.479$\pm$0.029 & 1.478$\pm$0.030 \\
\hline
nc - (nl+nr)/2 & 232$\pm$62 ppm & 237$\pm$64 ppm \\
\hline
\end{tabular}
\tablefoot{Choice of passbands and apertures are detailed in Section \ref{sec:hst_extraction}.}
\end{table}

\section{Simulations}
\label{sec:simulations}

In the following sections, we describe our simulated time series of spectra for the transit of HD209458~b, taking into account the stellar limb-darkening and rotation, for a variety of planetary atmosphere models. Then, we processed these synthetic data following typical procedures adopted in transmission spectroscopy studies at low and high resolution, and in particular the average spectra and light-curves presented as in \cite{charbonneau2002} and \cite{casasayas-barris2021}. The main purpose of this exercise is to investigate whether there are physical configurations that are compatible with the observations both at low and high resolution.
The methodology adopted to simulate the spectra in this paper is a generalisation of that described by \cite{casasayas-barris2019}. The main upgrade is related to the use of a wavelength-dependent planet radius to include the effect of its atmosphere.

\subsection{Modelling the star}
\label{sec:sim_star}
We modelled the sky-projected star disc as a grid of square cells, wherein each cell has an associated spectrum. Each cell is identified by the Cartesian coordinates of its centre $(x_j, \, y_j)$, assuming that the star disc is a circle with unit radius centred on the origin. The cell spectrum depends on its position because of the centre-to-limb variation (CLV, aka limb darkening) and the local radial velocity due to the stellar rotation.
We used the Spectroscopy Made Easy (SME, \citealp{valenti1996,piskunov2017}) package to generate intensity spectra for a star similar to HD209458 without rotational broadening. The SME calculation relies on a pre-computed grid of \texttt{MARCS} stellar atmosphere models \citep{gustafsson2008}. The spectral resolving power is $\sim$8$\times$10$^5$. These intensity spectra were calculated for 21 angles, equally spaced in $\mu$, except for $\mu =$0.005 instead of $\mu =$0 to avoid the singularity at the edge of the disc. Here, $\mu=\cos{\theta}$, where $\theta$ is the angle between the line of sight and the surface normal. The static cell spectrum is obtained by $\mu$-interpolation from the pre-calculated intensity spectra, where for a given cell
\begin{equation}
\mu_j = \sqrt{1 - x_j^2 - y_j^2} .
\end{equation}
In the star rest frame, the radial velocity of a given cell is
\begin{equation}
v_j = v \sin{i_*} \left ( y_j \sin{\lambda} - x_j \cos{\lambda} \right ) ,
\end{equation}
where $v \sin{i_*}$ is the radial component of the equatorial velocity, and $\lambda$ is the sky-projected obliquity.
We computed the Doppler shifted cell spectra in the star rest frame over the same wavelengths of the SME models. In this way, the disc-integrated stellar spectrum is the sum of the Doppler shifted cell spectra multiplied by the cell area.

\subsection{Modelling the planet}
\label{sec:sim_planet}
\begin{figure*}
\centering
\includegraphics[width=\hsize]{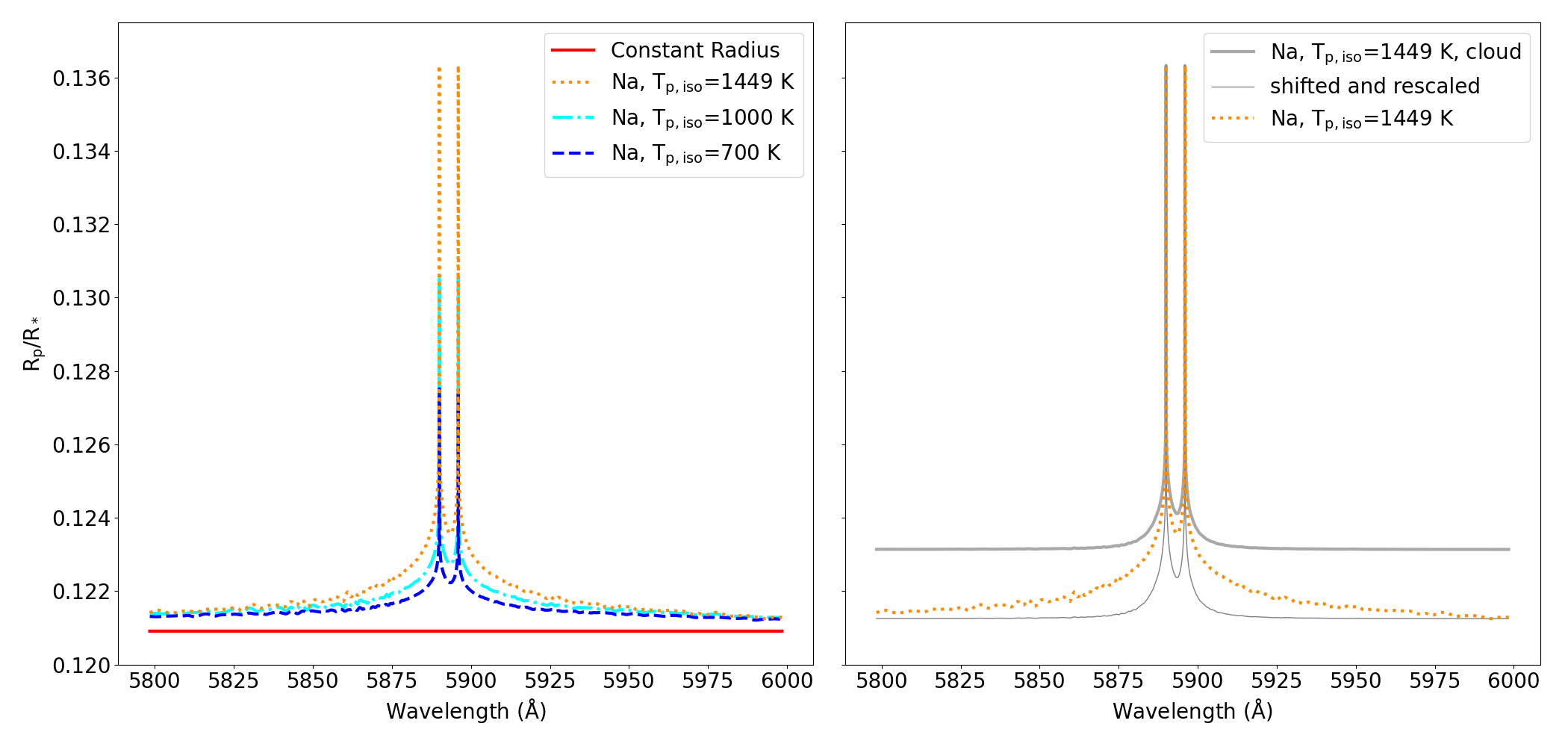}
\caption{Planet-to-star radii ratios versus wavelength. Left: constant radius (solid red line) and three pRT models with clear atmosphere and terminator temperatures of 1449 K (dotted orange), 1000 K (dash-dotted cyan), and 700 K (dashed blue). Right: pRT model assuming the equilibrium temperature at terminator with a cloud deck (solid grey thick line), shifted and rescaled model to have the continuum baseline and line peaks of the clear atmosphere model (solid grey thin line), and cloud-free model with the same temperature (dotted orange). 
}
\label{fig:Na_abs}
\end{figure*}

The planet is represented by a non-emitting opaque disc with a wavelength-dependent radius to account for atmospheric absorption. We modelled the case with constant radius equal to the value reported in Table \ref{tab:sys_params}, three cases with an atmosphere including \ion{Na}{i} absorption signature at different terminator temperatures, and another case that also includes clouds.
We used the petitRADTRANS\footnote{\url{https://gitlab.com/mauricemolli/petitRADTRANS/}} (pRT, \citealp{molliere2019}) package to compute the apparent planetary radii at multiple wavelengths including its atmosphere. The spectral resolving power was set to 10$^6$. Three of the simulated models assume clear atmospheres with solar abundances and a range of isothermal temperatures ($T_{\mathrm{p, iso}} = $ 700, 1000, and 1449 K). The fourth model, with $T_{\mathrm{p, iso}} = $ 1449 K, includes a grey cloud deck with top pressure of 1.38 mbar, as estimated by recent retrieval studies \citep{tsiaras2018}. The main effect of temperature is that it changes the strength of the \ion{Na}{i} lines, while the cloud deck also dampens their tails (see Figure \ref{fig:Na_abs}).
We note that these are the radii in the planet rest frame. Absorption from the planet atmosphere in the star rest frame is Doppler shifted due to the relative radial velocity of the planet,
\begin{equation}
V_{\mathrm{p}} (\phi(t)) = K_{\mathrm{p}} \sin{(2\pi \phi(t))} ,
\end{equation}
for a circular orbit. Here ,$K_{\mathrm{p}}$ is the orbital velocity of the planet and $\phi(t)$ is the orbital phase,
\begin{equation}
\phi(t) = \frac{t-E.T.}{P} ,
\end{equation}
with $E.T.$ being the reference epoch of mid-transit time, and $P$ being the orbital period.

\subsection{Modelling the transit}
\label{sec:sim_transit}
While transiting in front of its host star, the planet occults a time-varying portion of the stellar disc. In our simulations, we compute the unocculted spectrum at given instants with a cadence of 20 s. To do so, we first calculate the sky-projected planet coordinates $(X_{\mathrm{p}}(t), Y_{\mathrm{p}}(t))$ in units of the star radius using our own modified version of the \texttt{PYLIGHTCURVE} routine. Our version takes into account the light-travel delay \citep{kaplan2010,bloemen2012}, which has a negligible impact in this study. Second, we compute the relative planetary radii in the star rest frame, 
\begin{equation}
p( \lambda, t ) = \frac{ R_{\mathrm{p}} (\lambda, t)}{ R_* } ,
\end{equation}
where $R_{\mathrm{p}} (\lambda, t)$ are either constant or Doppler-shifted pRT models, interpolated over the same wavelengths of the cell spectra, and $R_*$ is the star radius.
Third, we apply the following mask to each cell spectrum:
\begin{equation}
m_j ( \lambda, t ) = \begin{cases}
0, & \text{if } dist( (X_\mathrm{p}(t), Y_\mathrm{p}(t)), \, (x_j, y_j) ) < p( \lambda, t ) \\
1, & \text{elsewhere}
\end{cases} ,\end{equation}
that is, the masked cell spectra are unchanged if the centre of the cell is external to the planet discs at all wavelengths, partially or fully zeroed otherwise. The stellar spectrum observed at a given instant is the sum of the cell spectra multiplied by the mask and by the cell area.

\begin{figure*}
\centering
\includegraphics[width=\textwidth]{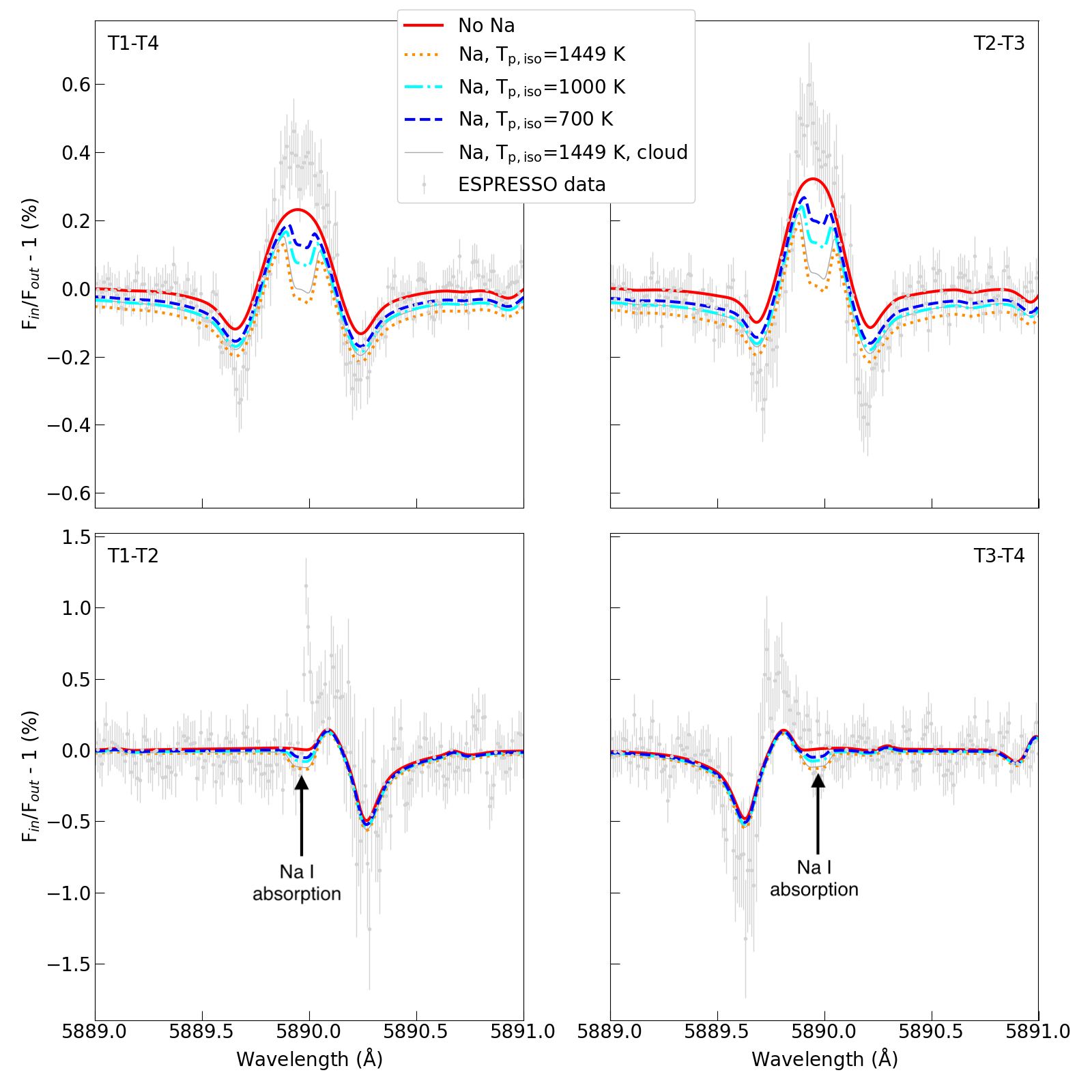}
\caption{Simulated transmission spectra of HD209458~b around the \ion{Na}{i} D1 line, showing the RM+CLV effects along with atmospheric absorption. The spectra are computed in the planet rest frame and averaged between transit contact points as indicated in the top left or top right corner of each panel. The different colours correspond to the same atmospheric models represented in Figure \ref{fig:Na_abs}. The ESPRESSO data are the same as those reported in Figure 5 of \cite{casasayas-barris2021}.
}
\label{fig:nuria_tspec}
\end{figure*}

\begin{figure*}
\centering
\includegraphics[width=\textwidth]{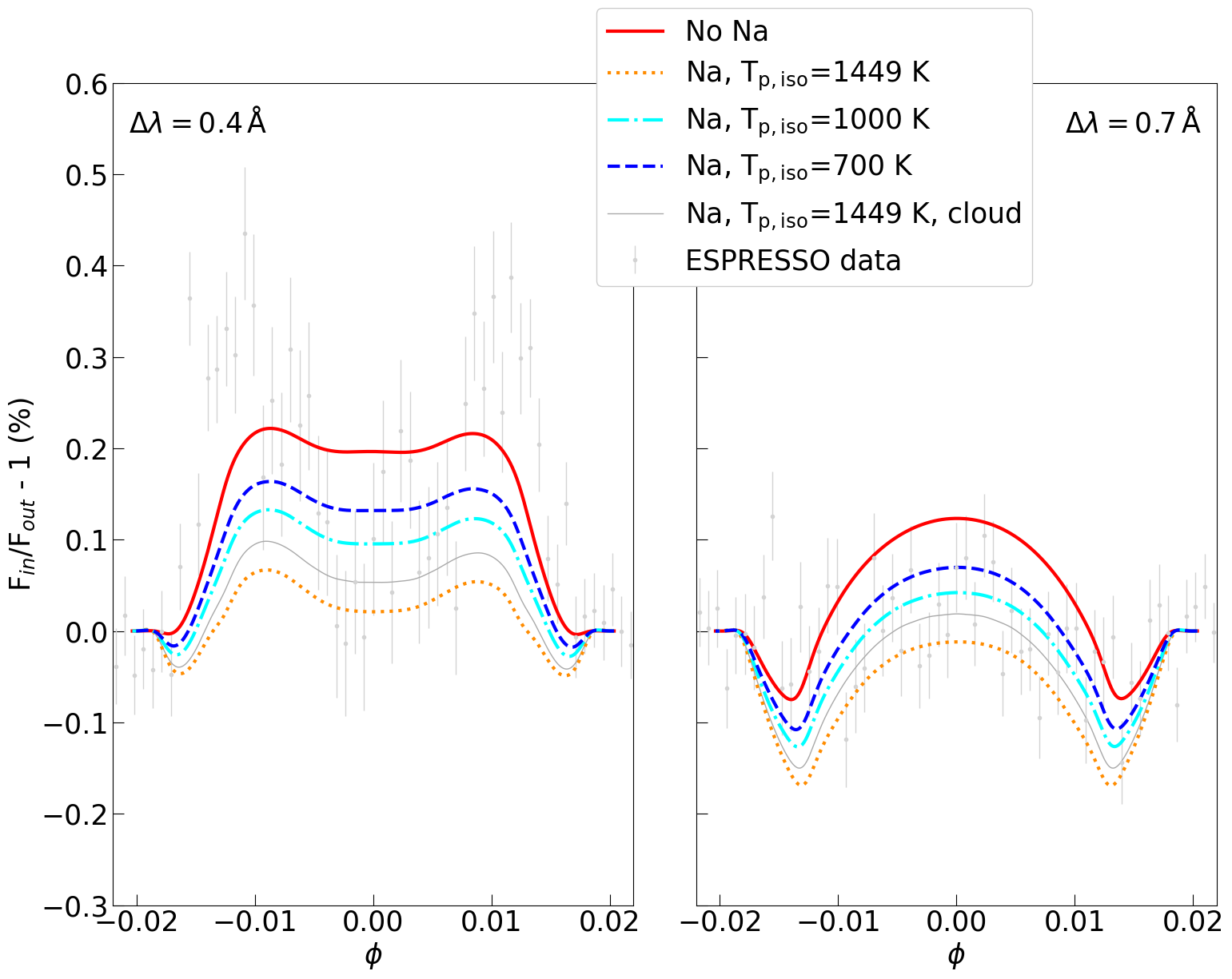}
\caption{Simulated transmission light curve of HD209458~b for passbands centred on any \ion{Na}{i} D line with $\Delta \lambda= \mathrm{0.4 \, \AA}$ and $\Delta \lambda= \mathrm{0.7 \, \AA}$, showing the RM+CLV effects along with atmospheric absorption. The light-curves are computed in the planet rest frame. The different colours correspond to the same atmospheric models represented in Figure \ref{fig:Na_abs}.  The ESPRESSO data are the same as reported in Figure 6 by \cite{casasayas-barris2021}.
}
\label{fig:nuria_lct}
\end{figure*}

\begin{figure*}
\centering
\includegraphics[width=0.95\textwidth]{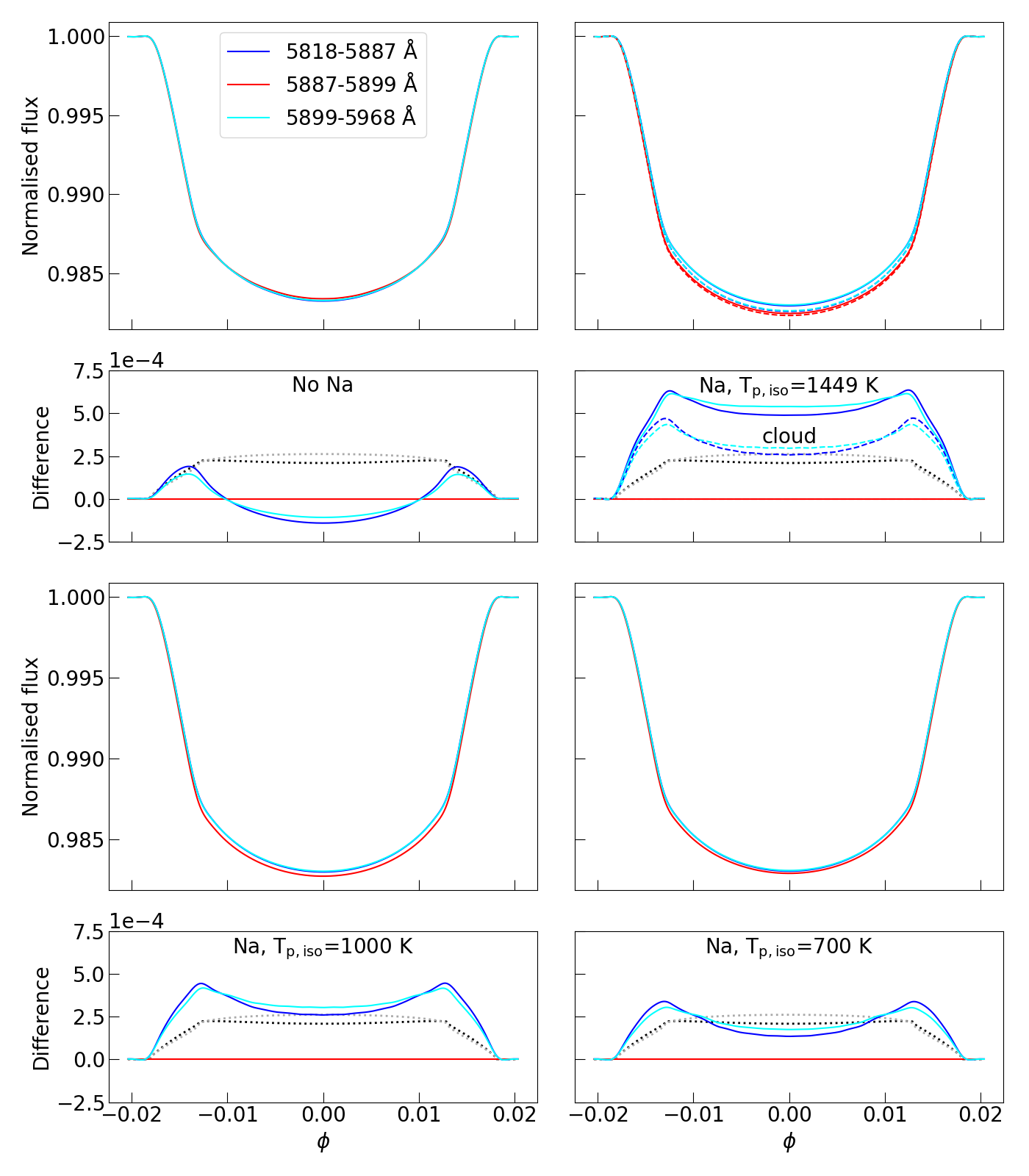}
\caption{Simulated transit light curves of HD209458~b for the same three passbands adopted in the HST data analysis and five planetary atmosphere models. The blue, red, and bright blue lines are obtained as described in Section \ref{sec:simulations}, with the continuous and dashed lines corresponding to clear and cloudy models, respectively. The bottom panels show the transit model differences with respect to that of the central band. The black and grey dotted lines are the HST model differences from Figure \ref{fig:tspec_cha02narrow}. We note that the HST model differences have a more trapezoidal shape than those obtained from the other set of simulations, because of the different limb-darkening profiles of the underlying \texttt{STAGGER} and \texttt{MARCS} stellar models.}
\label{fig:lc_charb02_simTcompar}
\end{figure*}

\subsection{Simulated transmission spectra and light curves}
\label{sec:sim_ts_lc}

Starting from the time series of simulated spectra, we calculated time-averaged spectra and band-integrated time series analogous to those that were used for data analysis and/or visualisation in previously published papers \citep{charbonneau2002,casasayas-barris2021}.
Figure \ref{fig:nuria_tspec} shows the transmission spectra in the planet rest frame averaged between different contact points, such as those reported in Figure 5 by \cite{casasayas-barris2021}. We zoomed in on the main \ion{Na}{i} line to highlight the effects of absorption in the planetary atmosphere, which are identical for both lines based on our models. The \ion{Na}{i} absorption imprints a dip near the peak of the symmetric RM feature obtained in the T1-T4 and T2-T3 intervals, and an overall offset due to its broad wings. The planetary \ion{Na}{i} may also appear as a small negative bump before or after the RM feature in the ingress or egress spectra.
We note that, depending on the S/N and resolving power of the spectroscopic data, the double-peaked RM feature in T1-T4 or T2-T3, and the small negative bumps in the T1-T2 and T3-T4 spectra may not be resolved. Also, the broadband spectral offset may be altered by some data-processing steps, in particular the continuum normalisation and the removal of instrumental systematic effects.
In these cases, the net effect of planet atmospheric absorption could be that of reducing the amplitude of the RM feature in T1-T4 and T2-T3 spectra. However, the feature amplitudes are also sensitive to the underlying stellar spectrum and limb-darkening profiles obtained with a different synthesis code and/or input physics \citep{casasayas-barris2021}. Consequently, the feature amplitude alone does not provide conclusive evidence about the presence or absence of \ion{Na}{i} absorption.

Figure \ref{fig:nuria_lct} shows the transmission light curves for bandwidths of 0.4 and 0.7 $\mathrm{\AA}$ centred on the \ion{Na}{i} D lines in the planet rest frame, as those reported in Figure 6 by \cite{casasayas-barris2021}. The \ion{Na}{i} absorption mostly affects the amplitude and vertical offset of the transmission time series relative to the out-of-transit baseline, and introduces only modest distortions during the ingress and egress phases. Even for these transmission light curves, the effects of \ion{Na}{i} in the planet atmosphere are degenerate with details of the stellar spectrum template.

Moving to low resolution, Figure \ref{fig:lc_charb02_simTcompar} shows the full transit light-curve models for the three HST passbands analysed in this study (Section \ref{sec:hst}) and by \cite{charbonneau2002}. Following the standard approach of low-resolution transit spectroscopy studies, they consist of instantaneous band-integrated fluxes relative to the stellar flux in the star rest frame. The decrease in normalised flux during transit corresponds to the fraction of stellar flux occulted by the whole planet disc, not just by its atmosphere. We subtracted the light-curve model of the central band to highlight the wavelength-dependent effects. If \ion{Na}{i} is not present in the planet atmosphere, the modulation obtained in the light-curve differences are solely due to the stellar RM and CLV effects. In this case, we measured a peak-to-peak amplitude of 330 ppm and mean value below 10 ppm for the modulation. If \ion{Na}{i} is present in the planet atmosphere, our models indicate positive light-curve differences at all phases, with median values of $\sim$190, $\sim$300, and $\sim$520 ppm assuming isothermal temperatures ($T_\mathrm{p, iso}$) of 700, 1000, and 1449 K, respectively. The cloudy model at 1449 K shows a median light-curve difference of 300 ppm.

We note that the light-curves plotted in Figures \ref{fig:nuria_lct} and \ref{fig:lc_charb02_simTcompar} have been smoothed to remove some fringing due to numerical errors, particularly in the models with high-resolution signatures. We checked that fringing was significantly reduced when decreasing the cell size adopted for the simulations, but the relevant calculations soon became too heavy. The smoothed light curves are a good compromise between precision of the models and computational costs.

\section{Discussion}
\label{sec:discussion}

\subsection{Comparison with published low-resolution spectroscopy}
\label{sec:lowres_compar}

\cite{charbonneau2002} announced the first detection of \ion{Na}{i} in the atmosphere of HD209458~b using the same HST data presented in Table \ref{tab:hst_obs_id}. They reported an absorption depth of 232$\pm$57 ppm based on the same passbands of Table \ref{tab:hst_tdepths}, but with a significantly different methodology. More specifically, they computed the absorption light curve as a linear combination of the three raw light curves, hence the depth is the difference between the out-of-transit and in-transit mean values. Instrumental systematics and stellar limb-darkening effects are ignored with their approach, although they were mitigated by removing the first point of each HST orbit and those during the transit ingress and egress.
In this work, we instead measured the transit depth for each passband by performing independent data detrending and light-curve fitting, and assuming a \texttt{STAGGER} stellar limb-darkening model (see Section \ref{sec:hst}). The absorption depth is a linear combination of the three transit depths. Our best result of 232$\pm$62 ppm matches that reported by \cite{charbonneau2002}.
Turning to empirical limb-darkening coefficients in the light-curve fits, however, we found that the HST data are consistent with the non-detection of \ion{Na}{i} in the atmosphere of HD209458~b.

By comparing the model light-curve differences shown in Figure \ref{fig:lc_charb02_simTcompar} with the best-fit results of Figure \ref{fig:tspec_cha02narrow}, the models that better reproduce the HST results are those with cloud-free atmosphere and terminator temperatures of 700-1000~K, or that with partially cloudy atmosphere and terminator temperature equal to the equilibrium temperature of 1449 K. We note that there is a degeneracy between the terminator temperature and cloud top pressure for a given amplitude of the \ion{Na}{i} absorption signal at low resolution \citep{lecavelier_des_etangs2008,benneke2013,heng2017}.

\cite{sing2008} analysed the transmission spectrum of HD209458~b with various resolving powers, obtaining similar results when considering the same passbands as \cite{charbonneau2002}. Additionally, \cite{sing2008} reported a spectral signal in the opposite direction to that of \ion{Na}{i} absorption at the line cores, that they attributed to telluric contamination. The removal of this effect would further increase the significance of the measured \ion{Na}{i} absorption. However, as pointed out by \cite{casasayas-barris2020}, a much more likely explanation for this signal is the combination of RM+CLV effects instead of a telluric origin.  

\subsection{Comparison with published high-resolution spectroscopy}
\label{sec:highres_compar}

\cite{snellen2008} and \cite{albrecht2009} analysed ground-based spectra taken with the High Dispersion Spectrograph (HDS) mounted on the Subaru telescope at Mauna Kea in Hawaii, and with the Ultraviolet and Visual Echelle Spectrograph (UVES) mounted on the Very Large Telescope (VLT) at Cerro Paranal in Chile. Their results are in good agreement with those reported by \cite{sing2008} based on HST data, when considering similar passbands centred on the \ion{Na}{i} D lines.

Recently, \cite{casasayas-barris2020} analysed the high-resolution spectra taken with TNG/HARPS-N, and with the Calar Alto high Resolution search for M dwarfs with Exoearths with Near-infrared and optical Echelle Spectrographs (CARMENES) located in Spain. Their approach differs from that of previous studies as the transmission spectra are computed in the planet rest frame instead of the stellar rest frame. They revealed the RM effect around the \ion{Na}{i} D, as well as other lines (H$\alpha$, \ion{Ca}{ii} IRT, \ion{Mg}{i} and \ion{K}{i} D1), without evidence of absorption in the exoplanet atmospheres. \cite{casasayas-barris2021} confirmed these findings with higher confidence and also reported an analogous behaviour for other lines (\ion{Fe}{i}, \ion{Fe}{ii}, \ion{Ca}{i,} and \ion{V}{i}), based on data taken with VLT/ESPRESSO. The ESPRESSO datasets have higher S/N than any other ground-based high-resolution dataset.

In Section \ref{sec:sim_ts_lc}, we discuss how \ion{Na}{i} absorption may alter the RM+CLV signal in transmission spectra. The error bars obtained by \cite{casasayas-barris2021} with ESPRESSO range from 0.05$\%$ to 0.09$\%$ for bin widths of 0.03 $\AA$. Based on the models plotted in Figure \ref{fig:nuria_tspec}, they should have resolved the double peak caused by \ion{Na}{i} absorption for both clear and cloudy atmospheric models with equilibrium temperature of 1449 K, if present, in the ESPRESSO data. We note that clouds tend to dampen the line tails, and hence they mostly affect the low-resolution signal. Models with cooler temperatures pose a greater challenge for the high-resolution signal, as the smaller dips on top of the RM feature can be comparable with the spectral error bars. If unresolved, the \ion{Na}{i} dip would decrease the amplitude of the RM feature.
The ESPRESSO data produces a signal of greater amplitude than that predicted by the models shown in Figure \ref{fig:nuria_tspec}, contrary to expectations in the presence of unresolved \ion{Na}{i} absorption.
However, the amplitude of the RM feature alone cannot be used to validate \ion{Na}{i} absorption in the atmosphere of HD209458~b, given the uncertainties associated with stellar models and system parameters. For example, \cite{casasayas-barris2021} obtained a larger RM feature by assuming a MARCS model without local thermodynamic equilibrium (see their Figure 10).
We also note that the broadband spectral offset is $\sim$0.05$\%$ for the model with the strongest absorption and it can be attenuated by ESPRESSO data processing, in particular due to continuum normalisation and removal of observed fringe patterns \citep{casasayas-barris2021}.

Similar considerations apply to the transmission light curves as shown in Figure \ref{fig:nuria_lct}. The possible \ion{Na}{i} absorption affects the amplitude and offset of the time signal, without significant distortions in most cases. The shape of the transmission light curves is strongly dependent on the CLV effect, for which the stellar models considered by \cite{casasayas-barris2021}, which are the same adopted in this work, do not provide a good match to the ESPRESSO data.

In conclusion, we find that atmospheric models with terminator temperature of 700-1000~K can be compatible with the non-detection of \ion{Na}{i} at high resolution. In Section \ref{sec:lowres_compar}, these models were also selected among the best fits to the low-resolution data that led to the 3.7$\sigma$ detection of \ion{Na}{i}. The high-resolution observations taken with ESPRESSO added further constraints; for example, discarding the hotter model with clouds.

\subsection{Planet atmospheric temperature}

\cite{morello2021} calculated an equilibrium temperature of 1449~K for HD209458~b in the case of zero reflectance and maximum circulation efficiency (see their Table 3). Taking into account the Spitzer phase-curve measured by \cite{zellem2014}, the terminator temperature should lie between the night-side temperature of 970 K and the day-side temperature of 1500 K. This temperature range can be further extended depending on the vertical profiles (e.g. \citealp{venot2019,drummond2020}). In principle, the 3D structure of the planetary atmosphere should be fully considered to simulate accurate transmission spectra, but 1D models can also reproduce the same spectra albeit with biases in physical and chemical parameters \citep{caldas2019,pluriel2020}. In particular, \cite{macdonald2020} demonstrated that the 1D retrieval techniques can lead to significantly underestimated temperatures for atmospheres with differing morning-evening terminators. This effect may explain the trend that terminator temperatures retrieved from transmission spectra are typically cooler than equilibrium temperatures \citep{fisher2018,tsiaras2018,pinhas2019}. Our isothermal temperature estimate of 700-1000 K for the atmosphere of HD209458~b aligns with this trend.

\section{Conclusions}
\label{sec:conclusions}

We investigated the puzzling question about the presence of \ion{Na}{i} in the atmosphere of HD209458~b, following conflicting reports from previous analyses of transit observations with low- and high-resolution spectroscopy. 
By comparing a set of models to the observations, we find that several atmospheric scenarios are compatible with both low- and high-resolution data. Overall, the HST and ESPRESSO datasets are consistent with a total absence of \ion{Na}{i} from the planetary atmosphere; otherwise, the terminator temperature of HD209458~b has to have an upper limit of about 1000K. More precise knowledge of the stellar intensity spectra, along with higher S/N data, is crucial to selecting the best scenario. The lower-than-equilibrium temperature (1449~K) on the terminator can be a consequence of heat redistribution processes or the effect of 1D model bias. Clouds may be present, in agreement with previous estimates, but they alone cannot help to reconcile the low- and high-resolution observations.
In some configurations, the presence of \ion{Na}{i} is clear from the broadband transmission spectrum, but, under certain physical conditions, the RM effect can mask the line cores at high resolution. 
If the overlapping signals at high resolution are not disentangled, the information on the spectral continuum from low-resolution observations can be decisive in detecting \ion{Na}{i} absorption.

This study highlights the complementarity between low- and high-resolution spectroscopic techniques for the characterisation of exoplanet systems.
In the literature, there are several contrasting results about exoplanet atmospheres based on low- and high-resolution spectroscopy \citep{huitson2013,sedaghati2017,espinoza2019,chen2018,allart2020}. We suggest that joint modelling efforts, such as the one proposed in this paper, could lift the apparent discrepancies. 
Considering a broader wavelength coverage should help reduce the degeneracy between stellar and planetary signals by considering multiple lines, therefore leading to tighter constraints on the possible exoplanet atmospheric models.

\begin{acknowledgements}
The authors would like to thank the referee, Jens Hoeijmakers, for useful comments that have improved the manuscript.
G. Mo. has received funding from the European Union's Horizon 2020 research and innovation programme under the Marie Sk\l{}odowska-Curie grant agreement No. 895525. N. C.-B. acknowledges funding from the European Research Council under the European Union's Horizon 2020 research and innovation program under grant agreement No. 694513. J. O.-M. and E. P. were financed by the Spanish Ministry of Economics and Competitiveness through grants PGC2018-098153-B-C31.
G. Mi. e G. C. acknowledge contribution from ASI-INAF agreement 2021-5-HH.0.
\end{acknowledgements}

%
%

\bibliographystyle{aa} 
\bibliography{biblio} 

\end{document}